\documentclass{article}
\usepackage{graphicx} 
\usepackage[margin=2.5cm]{geometry}
\usepackage{hyperref}
\usepackage{dsfont}

\usepackage{authblk}

\usepackage[english]{babel}

\usepackage{amsmath,amssymb}
\usepackage{xcolor}
\usepackage{hyperref}

\hypersetup{
    colorlinks=true,     
    linkcolor=black,         
    citecolor=black,       
    filecolor=black,    
    urlcolor=cyan
    }
    
\author{Carles Falc\'o\footnote{falcoigandia@maths.ox.ac.uk}$^{\hspace{.15cm}1}$, Ruth E. Baker\footnote{baker@maths.ox.ac.uk}$^{\hspace{.15cm}1}$, and Jos\'e A. Carrillo\footnote{carrillo@maths.ox.ac.uk}$^{\hspace{.15cm}1}$}
\date{}
\title{A nonlocal-to-local approach to aggregation-diffusion equations}

\usepackage{hyperref}

\hypersetup{
    colorlinks=true,     
    linkcolor=black,         
    citecolor=black,       
    filecolor=black,    
    urlcolor=cyan
    }

\begin{document}

\maketitle

\begin{center}
{$^{1}$\emph{Mathematical Institute, University of Oxford, Oxford, United Kingdom OX2 6GG }}
\end{center}

\begin{abstract}

Over the past decades, nonlocal models have been widely used to describe aggregation phenomena in biology, physics, engineering, and the social sciences. These are often derived as mean-field limits of attraction-repulsion agent-based models, and consist of systems of nonlocal partial differential equations. Using differential adhesion between cells as a biological case study, we introduce a novel local model of aggregation-diffusion phenomena. This system of local aggregation-diffusion equations is fourth-order, resembling thin-film or Cahn-Hilliard type equations. In this framework, cell sorting phenomena are explained through relative surface tensions between distinct cell types. The local model emerges as a limiting case of short-range interactions, providing a significant simplification of earlier nonlocal models, while preserving the same phenomenology. This simplification makes the model easier to implement numerically and more amenable to calibration to quantitative data. Additionally, we discuss recent analytical results based on the gradient-flow structure of the model, along with open problems and future research directions.

\end{abstract}

\emph{Keywords: }
aggregation-diffusion, cell-cell adhesion, thin-film equation, Cahn-Hilliard equation

\vfill
\newpage

\section{Introduction}
Agent-based models based on attraction-repulsion are ubiquitous across the sciences. From human crowd dynamics~\cite{bailo2018crowd} to cell sorting in developmental biology~\cite{VolkeningZebrafish}, a wide range of systems can be described using long-range attraction --- either towards a target or between agents --- coupled with short-range repulsion to prevent overcrowding or to model volume exclusion. Originally developed for studying many-body systems, such as gravitational and electrostatic interactions, this framework has found applications in modern fields like biology, social sciences, machine learning, and optimization. Attraction-repulsion models are typically formulated as systems of ordinary differential equations, and they become computationally expensive when the number of agents is very large, necessitating the use of macroscopic models. In the mean-field limit, attraction-repulsion dynamics are described by the aggregation-diffusion equation, a nonlocal partial differential equation. The study of aggregation-diffusion phenomena remains an active field, with ongoing advancements in the analysis, development, and application of mathematical models --- we refer to the recent surveys~\cite{carrillo2024aggregation, gomez2024beginner} for detailed explanations and additional references.

We begin by deriving one of the simplest mathematical models for attraction-repulsion dynamics, and explain its connection to the aggregation-diffusion partial differential equation. We follow  an energy minimization argument. Consider a system of $N$ agents with positions $\{\mathbf{y}_i\}_{i=1}^N$, where interactions are governed by an interaction potential $W^N(\mathbf{y})$. This potential is typically radially symmetric, and encodes both attractive and repulsive forces: it is large for short distances to prevent agents from overlapping (repulsion) and for long distances to maintain group cohesion (attraction). Then, the total potential energy for this system is given by $
E = \sum_{i=1}^N \sum_{j < i} W^N(\mathbf{y}_i - \mathbf{y}_j)
$. Assuming agents move to minimize $E$,
via gradient descent, their dynamics are given by 
\begin{equation*}
    \frac{\mathrm{d}\mathbf{y}_i}{\mathrm{d}t} = -\nabla_{\textbf{y}_i} E = -\sum_{j\neq i}\nabla W^N\left(\mathbf{y}_i-\mathbf{y}_j\right),\quad \mbox{for }i = 1,\ldots,N\,.
\end{equation*}
Note that the model above also corresponds to an overdamped version of Newton's second law, with forces given by $\nabla W^N$.

In the limit of large $N$, it is often more practical to describe the cell density, $\rho(\mathbf{x},t)$, for $\mathbf{x} \in \mathbb{R}^d$, rather than tracking the trajectories of individual agents. To do so, we define the empirical measure  
\begin{equation*}
\rho^N(\mathbf{x},t) = \frac{1}{N} \sum_{i=1}^N \delta_{\mathbf{y}_i(t)},
\end{equation*}
where $\delta_{\mathbf{y}_i(t)}$ represents the Dirac delta measure centered at the position $\mathbf{y}_i(t)$. Next, we consider the specific form of the potential $W^N$ and how it scales with the number of cells. To model volume exclusion, one can assume that for small distances, $W^N$ approaches a Dirac delta, $\delta_0$, at the origin as $N \to \infty$. This scaling has been rigorously analyzed in~\cite{Oelschlger1990LargeSO}, where the potential is written as  
\[
W^N(\mathbf{x}) = \epsilon N^{\beta - 1}\psi\left(N^{\beta/d}\mathbf{x}\right) + \frac{1}{N}W(\mathbf{x}),
\]  
with $\psi$ representing a typical repulsive potential with unit volume and $W$ a purely attractive potential. The parameter $\epsilon > 0$ quantifies the relative strength of repulsion to attraction.  Under this scaling, and for any $\beta \in (0,1)$, the empirical measure in the limit $N \to \infty$ can be shown to converge to the solution of the aggregation-diffusion equation
\begin{equation}
\label{eq:nonlocal_1species}
\frac{\partial\rho}{\partial t} = \nabla\cdot\left(\rho\nabla\left(\epsilon\rho + W * \rho\right)\right),
\end{equation}
where the convolution term is defined as $(W * \rho)(\mathbf{x},t)  = \int_\Omega W\left(\mathbf{x} - \mathbf{y}\right)\rho(\mathbf{y},t)\,\mathrm{d}\mathbf{y}$.

\subsection{Aggregation-diffusion in biology: modelling cell-cell adhesion}The above model is closely related to continuum models used to describe adhesion-based phenomena in biology~\cite{ArmstrongPainterSherratt,MurakawaTogashi}. These models capture how cells aggregate, spread, and form tissues through adhesion (attraction) and crowding constraints (repulsion). Cell-cell adhesion is a fundamental mechanism that regulates collective cell migration during tissue development, homeostasis, and repair. It allows cell populations to self-organize and ultimately form and maintain complex tissue structures. Cells interact with each other by forming protrusions, such as filopodia, and adhere to one another through the binding of cell surface proteins, and continuum models typically represent these interactions as nonlocal attractive forces~\cite{carrillo2024aggregation, buttenschon2020bridging, chenpainter2020nonlocal}.

Since Malcolm Steinberg formulated the differential adhesion hypothesis over 50 years ago~\cite{steinberg1962II,steinberg1962I,steinberg1962III,steinberg1963reconstruction}, numerous experimental and modeling efforts have explored the role of adhesion in cellular self-organization. The differential adhesion hypothesis proposes that differential adhesion between cell populations allows cells to sort and arrange themselves into complex patterns.  Steinberg developed a thermodynamic model in which groups of cells behave as immiscible fluids with different surface tensions, and where cells minimize the total adhesion energy by sorting themselves based on the relative strength of their adhesion bonds. This framework has successfully explained various tissue patterns, such as the bullseye formation, where two cell populations form concentric spheres with the more adhesive cells in the center~\cite{duguay2003cadherin,foty2005differential,krieg2008tensile}. The differential adhesion hypothesis continues to inspire both experimental and theoretical studies of cell sorting and tissue organization~\cite{foty2004cadherin, reviewCellCellAdhesion}.

Mathematical models describing  cell-cell adhesion have taken different approaches by considering either interfacial energy contributions, a tissue bulk modulus, or short-range attraction in the form of nonlocal interactions~\cite{AlertTrepat}. Many individual-based models have been used for adhesion-based patterning, e.g. cellular Potts type models~\cite{hirashima2017cellular,krieg2008tensile}, vertex models~\cite{VertexModels,hashimoto2017}, and particle-based models~\cite{CarrilloColombiScianna,VolkeningZebrafish}, to name but a few. While these have been successful in reproducing the observed experimental patterns, discrete models present difficulties, namely the computational cost involved in solving them and the lack of analytic insights. Continuum models can, in principle, address these issues, but their application to cell-cell adhesion  was only achieved nearly two decades ago.

The first continuum model capable of reproducing cell sorting phenomena was proposed by Armstrong et al.~\cite{ArmstrongPainterSherratt}. This model combines random motion, resulting in linear diffusion, with cell-cell adhesion, represented by a nonlocal attractive term. Cell-cell adhesion is modeled by assuming cells within a certain distance are attracted to each other. However, one limitation of this approach is that it fails to predict full segregation or sharp boundaries, as it assumes random motion. To overcome this, the linear diffusion term can be replaced with a density-dependent diffusion term that accounts for population pressure~\cite{MurakawaTogashi}. These nonlinear diffusion models are commonly used to describe crowding effects in mathematical biology~\cite{calvezCarrillo,DysonVolumeExclusion,Gurtin1977OnDiffusion} and have been derived from individual-based models~\cite{DysonMacroscopicCrowding} as well as on-lattice models~\cite{bakerAspectRation,falco2022random}. The modified model, along with variations incorporating density-limited mobilities~\cite{CarrilloMurakawaCellAdhesion}, has proven more effective in capturing the behavior of adhesion-based pattern formation.

The family of models with a nonlinear diffusion term modeling short-range repulsion is closely related to the model presented in Eq.~\eqref{eq:nonlocal_1species}. In the case of differential adhesion between two populations with densities $\rho$ and $\eta$~\cite{CarrilloMurakawaCellAdhesion}, we can apply similar ideas to derive the two-species model
\begin{subequations}
\begin{align}
    \frac{\partial \rho}{\partial t} &= \nabla\cdot\left(\rho\nabla \left(W_{11}*\rho + W_{12}*\eta + \epsilon(\rho + \eta) \right)\right), \label{eq:nonlocal_2species_a} \\
    \frac{\partial \eta}{\partial t} &= \nabla\cdot\left(\eta\nabla \left(W_{21}*\rho + W_{22}*\eta + \epsilon(\rho + \eta) \right)\right), \label{eq:nonlocal_2species_b}
\end{align}
\label{eq:nonlocal_2species}
\end{subequations}
 \hspace{-.19cm} where $W_{11}$ and $W_{22}$ represent the self-adhesion potentials, while $W_{12}$ and $W_{21}$ describe the cross-adhesion interactions. The parameter $\epsilon > 0$ quantifies the strength of the localized repulsion. The existence of solutions for this system is established in~\cite{AntonioTwoSpeciesNonlocal}, with further details in~\cite{difrancesco2013measure} for the case without cross-diffusion. A common assumption is that the cross-interaction is symmetric, i.e., $W_{12} = W_{21}$, and that the potentials share the same functional form: $W_{ij} = K_{ij} W$, where $W$ is a typical attractive potential and $K_{ij} \geq 0$ are constants representing the cell-cell adhesion strengths. This assumption is consistent with previous nonlocal models of cell-cell adhesion~\cite{ArmstrongPainterSherratt,CarrilloMurakawaCellAdhesion,MurakawaTogashi}. Although these models have been successful in understanding the underlying biology, they are mathematically complex, reducing analytical insight and limiting the extent to which they can be calibrated and used to describe experimental data.

\subsection{A local aggregation-diffusion model}

In this paper, following~\cite{falco2024local}, we present a local continuum model of aggregation-diffusion phenomena, where interactions between species are determined by surface tension parameters. Specifically, in the context of differential cell-cell adhesion, the local model reproduces the different patterns described by the differential adhesion hypothesis, in terms of differential surface tension, as originally proposed by Steinberg. This local model is derived from a general nonlocal model in the limit of short-range interactions and is expressed through a system of thin-film type equations. In its simplified form, the model has four parameters, each of which has a clear physical interpretation, both in the context of nonlocal models and as surface tensions in thin-film equations~\cite{myers1998thin}. Additionally, the local model offers the advantage of being more analytically tractable than nonlocal models, providing explicit stationary solutions even for two interacting species.

We follow the approach in~\cite{BernoffTopazCH} to derive a local model of cell-cell adhesion from Eqs.~\eqref{eq:nonlocal_2species}. However, the main focus of this paper is not to compare local and nonlocal models, but to study the local model and investigate its consistency with adhesion-based pattern formation and the differential adhesion hypothesis. By formally taking the limit of short-range interactions in the general nonlocal model described by Eqs.~\eqref{eq:nonlocal_2species}, we obtain a system of thin-film-like equations that model the evolution of the two cell populations
\begin{subequations}
\begin{align}
     \frac{\partial \rho}{\partial t}& = -\nabla\cdot\left(\rho\nabla \left(\kappa\Delta \rho + \alpha\Delta \eta + \mu\rho + \omega\eta \right)\right),\label{eq:intro1}\\
     \frac{\partial \eta}{\partial t}& = -\nabla\cdot\left(\eta\nabla \left(\alpha\Delta \rho + \Delta \eta + \omega\rho + \eta \right)\right)\label{eq:intro2}.
\end{align}
\label{eq:local_2species_intro}
\end{subequations} 
\hspace{-.21cm} The parameters in the system, $\kappa,\alpha,\mu\geq 0,\,\omega\in\mathbb{R}$, can be related to the potentials of the nonlocal model, $W_{ij}$, and to the strength of the volume-filling mechanism, but can also be understood as relative surface tensions, as in the thin-film equation. In this setting, one could ask whether differential tension --- analogous to differential adhesion --- in the model is sufficient to give rise to the  patterns seen in the Steinberg experiments. Interestingly, we show that it is possible to identify parameter regimes for each one of the different observed configurations (Figure \ref{fig:DAH_twospecies}) with the cross-interaction parameters $\alpha$ and $\omega$ playing a major role in the behavior of the local model (see Figure \ref{fig:4patterns}).

\subsection{Recent results} In the last years, the idea of approximating nonlocal models of biological aggregation with local equations has gained traction. This technique has been applied, for example, to determine the long-term behavior of metastable states in nonlocal models with linear diffusion~\cite{potts2024distinguishing}. More recently, Buttenschön et al.~\cite{buttenschon2024cells} used this approximation to demonstrate how cell-cell communication via chemotaxis can lead to stable cell clusters and numerically studied the nonlocal-to-local approximation for interaction kernels with exponential decay. Our work establishes a direct link between nonlocal aggregation-diffusion models and fourth-order equations of the Cahn-Hilliard or thin-film type. It extends the numerical explorations of Bernoff and Topaz~\cite{BernoffTopazCH}, who employed a truncated Fourier expansion of the interaction potential to derive a local model. Furthermore, analytical studies have since examined the nonlocal-to-local limit in both one-species~\cite{elbar2022degenerate} and two-species cases~\cite{carrillo2024degenerate}, providing rigorous justification for the limits used here in nonlocal models with compactly supported interaction potentials.

In recent work~\cite{carrillo2024competing}, we also develop the mathematical theory for a more general local model, establishing conditions for the global existence of weak solutions. The construction of these solutions relies on variational methods that exploit the gradient flow structure of the model, inherited from its nonlocal counterpart. Specifically, we prove the existence of weak solutions for a fourth-order equation with a more general aggregation term, 

\begin{equation*}
    \partial_t \rho = -\nabla \cdot (\rho \nabla (\Delta \rho)) - \chi \Delta \rho^m,
\end{equation*}
in the so-called subcritical regime $m < m_c = 2 + {2}/{d}$, as well as in the critical case $m = m_c$ for subcritical mass $\chi < \chi_c$ – analogously to Keller-Segel equations~\cite{calvezCarrillo}. Here, we derive the local model corresponding to the subcritical exponent $m = 2$. These results provide sharp conditions for global existence, leveraging uniform energy bounds and the minimising movement scheme. Moreover, we extend these results to a two-species system, proving the existence of weak solutions for coupled equations of the form of Eqs.~\eqref{eq:local_2species_intro}.

\subsection{Outline} This paper is structured in two parts. First we derive and study the local model for one cell population, including linear stability and numerical simulations of the model in one and two dimensions. Then, we extend these ideas and derive the model for two interacting cell populations, Eqs.~\eqref{eq:local_2species_intro}.
We show via numerical simulations that we can recover the patterns predicted by the differential adhesion hypothesis and interpret them in terms of model parameters.
While one of the advantages of the local model, compared to previously used nonlocal models, is the presence of analytically tractable steady states, we omit these calculations here for conciseness and refer to~\cite{falco2024local} for more details.
Finally, we summarize our findings and discuss  future research directions.

\section{One-species model}\label{sec:one_species}
\subsection{Heuristic derivation of the model and basic properties}

We begin with the nonlocal model given in Eq. ~\eqref{eq:nonlocal_1species}. Recall that the $\epsilon\rho$ term represents a localized repulsive force at the origin and the potential $W$ is assumed to be purely attractive and radially symmetric.

Many current models of adhesion only take into account interactions between cells that are separated by less than a maximum \emph{sensing radius}. Here, we build on the idea that for large populations, such \emph{sensing radius} is much smaller than the typical length of the population and hence attractive forces between cells are given by a short-range interaction potential. Hence, we set $W(\mathbf{x}) =- a^{-d}\varphi(\mathbf{x}/a)$ with $a$ a scaling parameter which dictates the range of attraction, and $\varphi$ a sufficiently smooth function defined in $\mathbb{R}^d$. As  $a\rightarrow 0$, the potential $W$ tends to a Dirac delta function supported at the origin.
We further assume that the function $\varphi$ satisfies several conditions:
\begin{enumerate}
    \item $\varphi(\mathbf{x}) = \varphi(|\mathbf{x}|)$ and $\varphi(r)$ is a non-increasing function for $r>0$, meaning that $W$ is both symmetric and attractive.
    \item $\varphi$ approaches a constant as $r\rightarrow\infty$. Without loss of generality we assume that this constant is zero.
    \item The moments   $M_{n} = \int_{\mathbb{R}^d}|\mathbf{x}|^n\varphi(\mathbf{x})\,\mathrm{d}\mathbf{x}$ decay suitably fast.

\end{enumerate}
Omitting the time dependence and using the change of variable \begin{equation*}
    (W*\rho)(\mathbf{x}) = -a^{-d}\int_{\mathbb{R}^d}\varphi\left(\frac{\mathbf{y}}{a}\right)\rho(\mathbf{x}-\mathbf{y})\mathrm{d}\mathbf{y} = -\int_{\mathbb{R}^d}\varphi(\mathbf{y})\rho(\mathbf{x}-a\mathbf{y})\mathrm{d}\mathbf{y},
\end{equation*} we can consider the limit of short-range attraction and expand $\rho(\mathbf{x}-a\mathbf{y})$ as a Taylor series for small values of the scaling parameter $a$:
\begin{align*}
     (W*\rho)(\mathbf{x}) =&  -\rho(\textbf{x})\int_{\mathbb{R}^d}\varphi(\textbf{y})\,\mathrm{d}\textbf{y}
          + a\int_{\mathbb{R}^d}\left(\nabla\rho(\textbf{x})\cdot \textbf{y}\right)\varphi(\textbf{y})\,\mathrm{d}\textbf{y}
    \\& -\frac{a^2}{2} \int_{\mathbb{R}^d}\left(\textbf{y}^T\cdot H_\rho(\textbf{x}) \textbf{y}\right)\varphi(\textbf{y})\,\mathrm{d}\textbf{y}+o(a^2);
\end{align*}
where $H_\rho(\textbf{x})$ is the Hessian matrix of $\rho$.

We will only keep the first terms in the expansion as these should give a good approximation for short-range interactions \cite{elbar2022degenerate}. For the first term in the Taylor expansion we simply have
    $\rho\int_{\mathbb{R}^d}\varphi = M_0\rho$, and we also note that the terms with odd order derivatives of $\rho$ vanish due to the symmetry assumption on the potential. Then the error term in the expression above is  $O(a^4)$.
    The next non-vanishing term in the series contains the second-order derivatives of $\rho$ and reads
    \begin{align*}
 \int_{\mathbb{R}^d}\left(\textbf{y}^t\cdot H_\rho(\textbf{x}) \textbf{y}\right)\varphi(\textbf{y})\,\mathrm{d}\textbf{y}&=\sum_{i = 1}^d\sum_{j = 1}^d\frac{\partial^2\rho}{\partial x_i\partial x_j}\int_{\mathbb{R}^d}y_iy_j\varphi(\mathbf{y}) \,\mathrm{d}\mathbf{y}\\& =   \sum_{i=1}^d\frac{\partial^2\rho}{\partial x_i^2}\int_{\mathbb{R}^d}y_i^2\varphi(\mathbf{y}) \,\mathrm{d}\mathbf{y}\\& = \frac{1}{d} \left(\int_{\mathbb{R}^d}|\mathbf{y}|^2\varphi(\mathbf{y}) \,\mathrm{d}\mathbf{y}\right)\sum_{i=1}^d\frac{\partial^2\rho}{\partial x_i^2}=\frac{M_2}{d}\Delta \rho,
 \end{align*}
where we used again that $\varphi$ is symmetric. Putting this all together gives
\begin{equation*}
    W*\rho = - M_0\rho -\frac{M_2 a^2}{2d}\Delta\rho +O(a^4M_4).
\end{equation*}
Using only the first two terms in the approximation in Eq.~\eqref{eq:nonlocal_1species} yields
\begin{equation}
      \frac{\partial\rho}{\partial t} = -\nabla\cdot\left(\rho\nabla \left(\tilde{M}\Delta \rho + (M_0-\epsilon)\rho \right)\right),
      \label{approximatedPDE}
\end{equation}
with $\tilde{M} = M_2 a^2/2d$. Note that the approximation makes sense as long as moments $M_n$ of higher order ($n\geq4$) are small compared to $M_2$, and the scaling parameter $a$ is small. We emphasize here though, that the goal of our paper is not to compare~\eqref{approximatedPDE} with the nonlocal model~\eqref{eq:nonlocal_1species}, but to study possible behaviors of the local model in Eq.~\eqref{approximatedPDE}. We refer to~\cite{carrillo2024degenerate, elbar2022degenerate} for rigorous justification of this limit.

Two relevant observations can be made here. First, note that the sign of $M_0-\epsilon$ gives the relative strength of repulsive and attractive forces. For negative $M_0-\epsilon$, localized repulsion is the dominant interaction, while for positive values of $M_0 - \epsilon$, the short-range attractive forces overcome repulsion. Here, we focus on the latter case, since it allows biological aggregation. In fact, with our choice of diffusion and aggregation potential $W$, Eq.~\eqref{eq:nonlocal_1species} only has stationary states in the $M_0-\epsilon > 0$ case~\cite{BurgerFranekSStates}. As we will see, our analysis suggests that this is also the case for the local model given by Eq.~\eqref{approximatedPDE}.

\begin{figure}
    \centerline{\includegraphics[width = \textwidth]{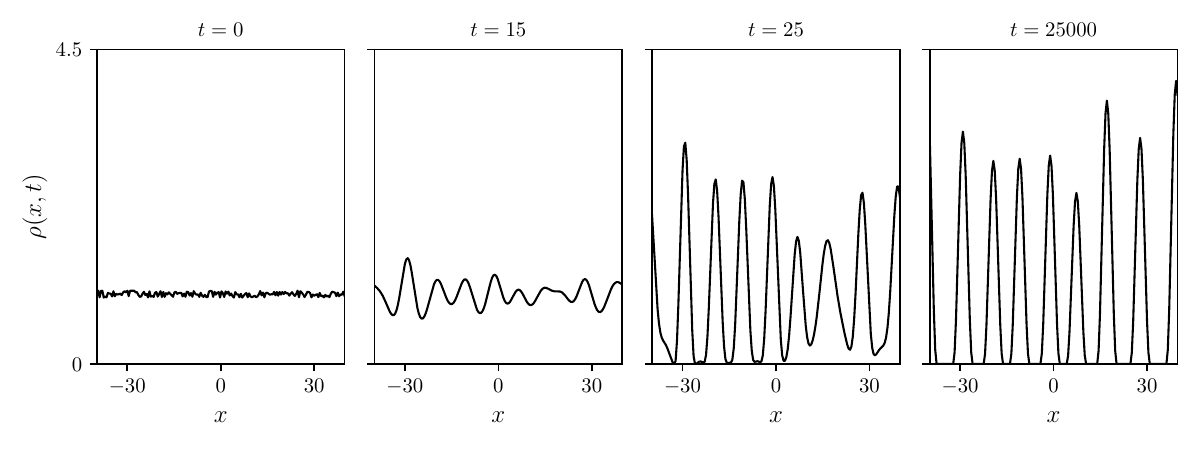}}
    \caption{Aggregation is possible in the local model as long as $\mu^2>0$. Numerical simulations with periodic boundary conditions and parameters: $\mu^2 = 1,\,L = 40,\,\Delta x = 0.2,\,\Delta t = 0.01$. Initial data corresponds to the spatially homogeneous steady state $\rho(x,0) = 1$ for $x\in[-L,L]$ plus a small perturbation.}\label{Fig:Aggregations1D}
\end{figure}

Secondly, and as it was already remarked in~\cite{ArmstrongPainterSherratt}, the fourth order term has a dampening effect on the PDE. In the absence of this term, one obtains an ill-posed problem due to the negative diffusion coefficient. Therefore, it does not seem possible to have a second-order model of cell-cell adhesion, thus making evident the need for a fourth-order approximation. A similar phenomenon happens in~\cite{negativeDiffusionAnguige} when one takes the continuum limit of a microscopic model incorporating cell-cell adhesion.

Before moving onto further considerations, and in order to facilitate the analysis, we nondimensionalize Eq.~\eqref{approximatedPDE}. Under a suitable rescaling --- for instance, set $\rho\mapsto \tilde{M}\rho$ and $\mu^2 = (M_0-\epsilon)/\tilde{M}$ --- the model can be written as
\begin{equation}
    \frac{\partial\rho}{\partial t} = -\nabla\cdot\left(\rho\nabla \left(\Delta \rho + \mu^2\rho \right)\right),
    \label{approximatedPDE_rescaled}
\end{equation}
with $\mu^2 > 0$, according to our previous considerations. This model resembles a Cahn-Hilliard~\cite{elliott1996cahn} or thin-film type equation where the parameter $\mu^{-2}$ plays the role of surface tension~\cite{myers1998thin}, which reintroduces the idea of modeling tissues as having fluid-like properties, as originally proposed by Steinberg in his model. These considerations will become more relevant later on when we discuss systems of two species.

The thin-film equation governs the evolution of the thickness of a thin fluid layer on a surface. Equations of the type of Eq.~\eqref{approximatedPDE_rescaled} appear as the lubrication approximation of a gravity-driven Hele-Shaw cell~\cite{HeleShaw2,goldstein1998instabilities}. Depending on the sign of $\mu^2$, the equation is referred as \emph{long-wave unstable} or \emph{long-wave stable}. The sign of $\mu^2$ characterizes the linear stability of the constant steady state --- more details will be discussed later. This model also falls under a larger family of thin-film equations, whose properties have been well-studied --- see~\cite{linearstabilitypugh,LAUGESEN2002377,laugesenpugh} for an exhaustive study of the steady states,~\cite{slepvcev2009linear} for stability of self-similar solutions, and ~\cite{bertozzi2,bertozzi1998long} for long-time behavior of solutions and regularity.

Associated with the local model, we also have the free energy
\begin{equation}
    \mathcal{F}[\rho]= \frac{1}{2}\int_\Omega\left(|\nabla\rho|^2 -\mu^2\rho^2\right) \mathrm{d}\mathbf{x}.
    \label{energyCH}
\end{equation}
With this in mind,~\eqref{approximatedPDE_rescaled} can be written as a gradient flow with respect to the 2-Wasserstein metric (see for instance~\cite{carrillo2003kinetic,matthes2009family,santambrogio2015optimal})
\begin{equation*}
    \frac{\partial\rho}{\partial t} = \nabla\cdot\left(\rho\,\nabla \frac{\delta \mathcal{F}}{\delta\rho}\right).
\end{equation*}
Observe that by integrating by parts formally, 
\begin{align*}
    \frac{\mathrm{d}}{\mathrm{d}t}\mathcal{F}[\rho] &= \int_\Omega \nabla\rho\cdot \nabla\left(\frac{\partial\rho}{\partial t}\right)\mathrm{d}\mathbf{x} - \mu^2\int_\Omega\rho\,\frac{\partial\rho}{\partial t}\,\mathrm{d}\mathbf{x}\\&=\int_\Omega\rho\,\nabla\left(\Delta\rho + \mu^2\rho\right)\cdot\nabla\left(\frac{\delta\mathcal{F}}{\delta\rho}\right)\mathrm{d}\mathbf{x} \\&=-\int_\Omega\rho\left|\nabla\frac{\delta\mathcal{F}}{\delta\rho}\right|^2\mathrm{d}\mathbf{x}\leq 0\,,
\end{align*}
and hence the energy is non-decreasing in time. Here we used the first and third order boundary conditions
\begin{equation}
    \label{eq:boundary_conditions}
    \partial_\nu\rho = \partial_\nu\Delta \rho = 0\quad\mbox{on }\partial\Omega,
\end{equation}
where $\nu$ is the exterior normal of $\Omega$.

\subsection{Linear stability analysis}
\label{section:linear_stability_analyis}
Equation~\eqref{approximatedPDE_rescaled} admits as steady states any spatially homogeneous solution. One of the first biologically relevant questions that arises from this model, is whether aggregations are possible as in the case of Eq.~\eqref{eq:nonlocal_1species}. To investigate this question we perform linear stability analysis on the spatially homogeneous solution $\rho(\textbf{x},t) = \rho_0$. In order to do so, we consider a perturbation $\rho(\textbf{x},t) = \rho_0 + \Tilde{\rho}(\textbf{x},t)$ and linearize the resulting equation. By setting $\Tilde{\rho}(\textbf{x},t)\propto e^{i\textbf{k}\cdot \textbf{x}+\sigma(\textbf{k})t}$ one finds the dispersion relation
\begin{equation*}
    \sigma(\textbf{k}) = \rho_0|\textbf{k}|^2\left(\mu^2-|\textbf{k}|^2\right).
\end{equation*}
In fact, the resulting linearized equation is identical to that of the standard Cahn-Hilliard equation ~\cite{elliottCH}.
A necessary condition for the formation of non-trivial stationary states is then $\text{Re}(\sigma(\textbf{k})) > 0$ for certain values of the wave vector $\textbf{k}$, which results in the upper bound: $|\textbf{k}| < \mu$.

 From the unstability condition we already see that in the case where $\mu^2<0$, the homogeneous steady state is linearly stable and thus aggregation is not possible. As mentioned earlier, this case happens when $-\int_{\mathbb{R}^d}W < \epsilon$, for which Eq.~\eqref{eq:nonlocal_1species} has no stationary states either~\cite{BurgerFranekSStates}. We note that these considerations are not new, as Eq.~\eqref{approximatedPDE_rescaled} belongs to a broader class of thin-film equations, whose linear stability is well-established~\cite{laugesenpugh}. From now on, we always consider the local model in the long-wave unstable regime $\mu^2>0$.

\subsection{Numerical experiments}
In this section, we explore numerically some basic properties of the local model. For that purpose we use a numerical scheme based on that in~\cite{bailo2021unconditional} (see~\cite{falco2024local} for more details). Moreover, we run all of our simulations on a domain $[-L,L]^d$, where $L$ and $d = 1,2$ are specified individually for every experiment. We also assume periodic boundary conditions on $\rho$ and its derivatives. We have also implemented the local model in VisualPDE~\cite{visualPDE}, providing a platform for rapid and interactive visualization of the model dynamics --- the animated simulation can be accessed \href{https://visualpde.com/sim/?options=N4IgRghiBcIM4FsD2SAuALCATLIA04AbjCAHQCM+IAxidegKYICW1EANgELsCuDV1ACIxUAJz4EsAVSkkAtAh4B9AEwAqBBAAeACh54ADAEoq0gGrzNu-cdNnZsSpLMXYB0wHUPJd5LgkVXxAsVBgDUgNIg3ICagcQJxAAaxJFVQACAF508nTmADt09IBtOQBWPCKygF0AbipNEgAWBuYfKnyoaBiQfIAVGHIoggAHEgBhHjhUJAQqEdEkMAYAMR58kh55xeWAGQZ8gHMMQYqQUXig0VcQK+83KjgGVABlBgZcaDEJeDG3UhUZUeG1gPHSxAIcAAjiJxAxIVsvnCCKhmAgGHA2OwMbCfsRoMVQBBRIsAO7jJDsJA8UQwABsZTKAGYzsSyYIDnBmKgAJ5hUisklIUn7I4YACy2hIiTZwpeWP4sAgPBmVFlpIAGiRdkotAB6FTpOTpLRqoWkgCa2qUPINRvSfIIypmb1QFKpNIAShAjoqAGYcJ6xCDo0QQPqMVBdJkGYOhiAABXQbWgMbjDDDAC0UHNoOEAByxSnU0SaP40RgsLHcPgCYs0yC06AB9hBmhIfIzGn+ZuB+Htzsl90lsJFwc0gCiIy5VJB4WiY67ogAcjxc2dqFMZggXjSA9R-X2CEwwEg4D2W22T2e4ABBBBgZgHUJ5gXHh83wTMP1+qaK8J0u+p7nkmKbkAKADsBiFiA17ni8yb5Hy3SxrBH7wcgaDoPkGI9uEKhATeLwjAwm7sMS-IEWhwFwBGzxdOEEH5mUTQEH67DMCMJFYMONJljAl79hxOEeMwIToOKPDsIMBBWLx1wcLWsAtLJBTyWYimKrcVBIIQGbkchgkELp+kQDy8mjiAJmiAZU4zh2-KRGc1m2VoCzSjpek2WZuwFAwoniZJ0ndKMVKoH0PIkSQIzkTh2xLAUhwCUeIALEsqzrJs8V7AcxzoKcow7AwEVRbABSoAwhxhtJhXpVqDy1csVoNfAe4QAeawgtpKLoKI7yCAAEgwzCHOgrwKvyEEEHp1AzKIKxPuwnxGSApKYOFSAJmFWUEJ06IeQAvtUBCkp8IBbCd5YXfAaJ9Ny2LakgWLpOMDDsOwcgHu96TYIwXIdukyBYG99q5HAJHUE+-gHUAA}{here}. We note that this implementation may introduce numerical errors, so we recommend using the finite-volume scheme in~\cite{falco2024local} for accurate results.

\begin{figure}
    \centerline{\includegraphics[width = \textwidth]{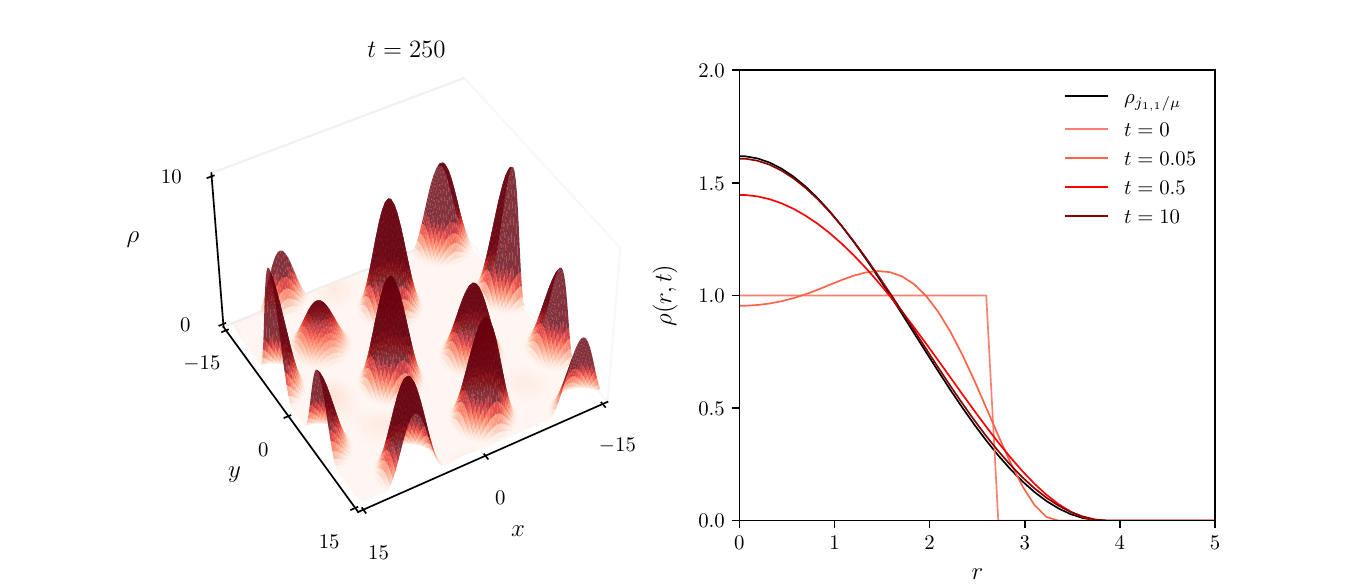}}
    \caption{(left) Aggregation in the two-dimensional local model. Initial data is $\rho(x,0) = 1$ plus a small perturbation. Density configuration at $t = 250$. (right) Convergence to steady state in the two-dimensional model. Radial density profiles at different time points and stationary solution given by $\rho_{j_{1,1}/\mu}(r)$ (see~\cite{falco2024local} for details on steady states). Simulation parameters are $\Delta t = 0.01,\,\Delta x=0.1,\,\Delta y =0.1,\,\mu = 1$ and domain specifications: $L = 15$ for \textup{(a)} and $L=5$ for \textup{(b)}.}\label{Fig:Aggregations2D}
\end{figure}

We start by validating some of the analytic results in the previous section. As expected from our derivations, the spatially homogeneous steady state is unstable in the case $\mu^2 > 0$. To test this prediction, we use a suitably large domain, $L = 40$, and perform simulations using as initial densities a slightly perturbed homogeneous steady state (Figure \ref{Fig:Aggregations1D}). We see that small perturbations rapidly lead to spatial patterning that mimic previous models of cell-cell adhesion~\cite{ArmstrongPainterSherratt}. While cell densities change very rapidly at early times, as they evolve towards different peaks, smaller density bumps disappear at a very low rate, only reaching the stationary configuration after much longer times. In particular, the local model shows similar behavior to nonlocal models with compactly supported interaction potentials. These typically give rise to stationary states with multiple separated aggregates~\cite{nonlocalSchemeCarrilloYanghong} where the distance separating different cell aggregates is larger than the \emph{sensing radius} in the potential. A similar pattern appears in the two-dimensional local model (Figure \ref{Fig:Aggregations2D}).

\section{Extension to two interacting populations}

Having analyzed the model for a single population, we now extend it to two interacting cell populations. Our main goal here is to study if such model is able to reproduce the patterns seen in the Steinberg experiments, and whether this behavior can be understood in terms of the model parameters.

\begin{figure}
    \centerline{\includegraphics[width = .95\textwidth]{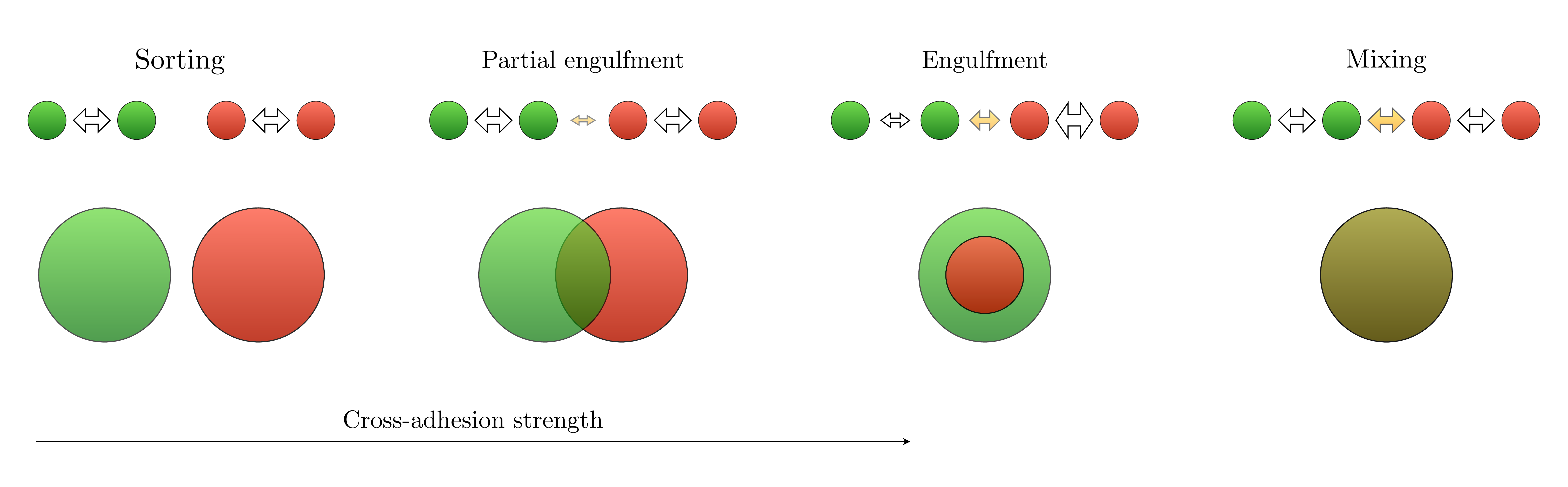}}
    \caption{Possible configurations for Steinberg experiments in terms of the cross-adhesion and the self-adhesion of a system of two species (adapted from \textup{\cite{MurakawaTogashi}}). In the weak cross-adhesion regime we might have two patterns depending on whether the cross-adhesion strength is strictly zero or positive. \textbf{Sorting} is observed when there is no cross-adhesion between the two species, and \textbf{partial engulfment} when cross-adhesion is small compared to the self-adhesion of each population. When the cross-adhesion is stronger, the system might evolve to an \textbf{engulfment} pattern, where the more cohesive species is surrounded by the less cohesive one; or to complete \textbf{mixing} of the cell populations. The first corresponds to the case in which the cross-adhesion is stronger than the self-adhesion of one species but weaker than the self-adhesion of the other one. The latter occurs when the cross-adhesion strength is comparable to both self-adhesion forces.}
    \label{fig:DAH_twospecies}
\end{figure}

In the case of two interacting populations, the self-adhesion of each species and the cross-adhesion between them determine the behavior of the system. Depending on the relative strength of adhesive forces, experiments show that the two cell populations can evolve into one of four distinct configurations, as depicted in Figure \ref{fig:DAH_twospecies}.

\subsection{A system of thin-film equations to model cell-cell adhesion}

We proceed as in the one-species case and assume that the potentials $W_{ij}$  in Eqs.~\eqref{eq:nonlocal_2species}  are attractive and scale according to a parameter $a$, which gives the range of interactions. More precisely we impose $W_{ij}(\textbf{x})=-a^{-d}\varphi_{ij}\left(\textbf{x}/a\right)$, with the functions $\varphi_{ij}$ satisfying the conditions in the previous section. In the limit $a\rightarrow 0$, we can approximate $W_{ij}*f\approx - c_{ij}f -d_{ij}\Delta f$, where $f \in\{ \rho,\eta\}$ and the constants $c_{ij},d_{ij}$ could be different for each potential. Note that $c_{ij}$ is the volume of $\varphi_{ij}$ and $d_{ij}$ is related to its second moment \begin{equation*}
    c_{ij} = \int_{\mathbb{R}^d} \varphi_{ij}(\textbf{x})\,\mathrm{d}\textbf{x},\quad d_{ij} = \frac{a^2}{2d}\int_{\mathbb{R}^d} |\textbf{x}|^2\varphi_{ij}(\textbf{x})\,\mathrm{d}\textbf{x}.
\end{equation*} For simplicity we assume here that the cross-interaction potential is the same for the two species $W_{12}=W_{21}$, which is a commonly used assumption in many models of cell-cell adhesion~\cite{ArmstrongPainterSherratt,CarrilloMurakawaCellAdhesion}. Using these approximations in the two species nonlocal model Eqs.~\eqref{eq:nonlocal_2species}, yields
\begin{subequations}
\begin{align*}
     \frac{\partial \rho}{\partial t} = -\nabla\cdot\left(\rho\nabla \left(\kappa_1\Delta \rho + \tilde{\alpha}\Delta \eta + \mu_1\rho + \tilde{\omega}\eta \right)\right),
     \\
     \frac{\partial \eta}{\partial t} = -\nabla\cdot\left(\eta\nabla \left(\tilde{\alpha}\Delta \rho + \kappa_2\Delta \eta + \tilde{\omega}\rho + \mu_2\eta \right)\right).
\end{align*}
\end{subequations}

The model parameters can be understood in terms of the potentials $W_{ij}$. First note that the parameters in the fourth order terms, $\kappa_1,\kappa_2$ and $\tilde{\alpha}$, are directly related to the second moments of the potentials. Hence, they only give information on the strength and range of attractive forces. Assuming that the potentials are all attractive, we have $\kappa_1,\kappa_2,\alpha\geq 0$. On the other hand, the parameters in the second order terms, $\mu_1,\mu_2$ and $\tilde{\omega}$, are both related to the volumes of the potentials, and to the strength of repulsive forces, which are given by $\epsilon$. According to our considerations in section \ref{section:linear_stability_analyis}, we assume that $-\int_{\mathbb{R}^d}W_{11} > \epsilon$ and also $-\int_{\mathbb{R}^d}W_{22} > \epsilon$, meaning that self-attraction overcomes repulsion in each  of the  populations. This gives $\mu_1,\mu_2>0$. However, cross-attraction between the two type of cell types could be weaker and thus $\tilde{\omega}$ could be either positive or negative.

In fact, and in order to facilitate the analysis, we can reduce the number of parameters with a suitable rescaling of the variables. For example, set $x \mapsto \xi x$, $t \mapsto T t$, $\rho\mapsto\mu_2\rho$, $\eta\mapsto\mu_2\eta$, with $\xi^2 = {\mu_2}/{\kappa_2}$ and $T = \xi^{2}$. This yields the new rescaled parameters
\begin{equation*}
     \kappa = \frac{\kappa_1}{\kappa_2}\,,\quad\alpha = \frac{\tilde{\alpha}}{\kappa_2}\,,\quad\mu = \frac{\mu_1}{\mu_2}\,,\quad\omega = \frac{\tilde{\omega}}{\mu_2}\,,
\end{equation*}
and the reduced system
\begin{subequations}
\begin{align}
     \frac{\partial \rho}{\partial t}& = -\nabla\cdot\left(\rho\nabla \left(\kappa\Delta \rho + \alpha\Delta \eta + \mu\rho + \omega\eta \right)\right),\label{eq:local_2species_a}\\
     \frac{\partial \eta}{\partial t}& = -\nabla\cdot\left(\eta\nabla \left(\alpha\Delta \rho + \Delta \eta + \omega\rho + \eta \right)\right)\label{eq:local_2species_b}.
\end{align}\label{eq:local_2species}
\end{subequations}
Here, $\kappa\geq 0$ and $\mu> 0$ represent the relative self-adhesion strength of $\rho$ with respect to $\eta$; while $\alpha\geq 0$ and $\omega\in\mathbb{R}$ give the relative strength of the cross-attraction forces.

Observe too that the local model, Eqs.~\eqref{eq:local_2species}, is essentially a system of two thin-film like equations, where the parameters $\kappa$ and $\alpha$ can be understood as the relative tensions of one species with respect to the other one, and of the interface separating the two populations. The parameters in the second order terms $\mu$ and $\omega$ are then related to the population pressure exerted by each cell type. Under this setting, it makes sense to ask whether differential tension, as originally proposed by Steinberg, can explain cell sorting phenomena. In other words, can we identify relevant regimes for the four parameters $\kappa,\alpha,\mu,\omega$ such that the experimental patterns are recovered in the local model? We highlight that the local model given in Eqs.~\eqref{eq:local_2species} is not a phenomenological description emerging from the tissue-fluid analogy, but arises in the limit of short-range interactions of Eqs.~\eqref{eq:nonlocal_2species}, and hence it provides a direct connection between the original differential adhesion hypothesis and a physical model of cell-cell adhesion.

In order to avoid negative diffusion, we require the matrix
\begin{equation*}
    M = \begin{pmatrix}
 \kappa & \alpha\\
 \alpha & 1\\
\end{pmatrix},\label{eq:M}
\end{equation*}
to be positive definite. Since $\kappa > 0$, this requires $\det M \geq 0$. This sets a limit on the strength of the cross-attraction: $0\leq \alpha<\sqrt{\kappa}$.
\subsection{Energy}

Thanks to the symmetry in the cross-interaction terms given by $\alpha$ and $\omega$, this system also exhibits a gradient-flow structure
\begin{subequations}
\begin{align*}
     \frac{\partial\rho}{\partial t} = \nabla\cdot\left(\rho\,\nabla \frac{\delta \mathcal{F}_2}{\delta\rho}\right),\\
         \frac{\partial\eta}{\partial t} = \nabla\cdot\left(\eta\,\nabla \frac{\delta \mathcal{F}_2}{\delta\eta}\right),
\end{align*}
\end{subequations}
with respect to the  2-Wasserstein metric~\cite{carrillo2024competing,carrillo2003kinetic,matthes2009family,santambrogio2015optimal} and the free energy
\begin{equation}
    \mathcal{F}_2[\rho,\eta] = \int_\Omega \left(\frac{\kappa}{2}|\nabla\rho|^2+\frac{1}{2}|\nabla\eta|^2+\alpha\nabla\rho\cdot\nabla\eta -\frac{\mu}{2}\rho^2-\frac{1}{2}\eta^2-\omega\rho\eta\right) \mathrm{d}\mathbf{x}.\label{eq:energy2}
\end{equation}
We note that the nonlocal model for two species, given by Eqs.~\eqref{eq:nonlocal_2species}, also exhibits a gradient flow structure when the cross-interaction potentials are symmetrizable, providing variational schemes to prove the existence of solutions to the system~\cite{difrancesco2013measure,AntonioTwoSpeciesNonlocal}.

As in the one-species case, we can formally integrate by parts to show that the energy is non-increasing in time
\begin{align*}
     \frac{\mathrm{d}}{\mathrm{d}t}\mathcal{F}_2[\rho,\eta]  =&\,\kappa\int_\Omega \nabla\rho\cdot \nabla\left(\frac{\partial\rho}{\partial t}\right)\mathrm{d}\mathbf{x} +   \int_\Omega \nabla\eta\cdot \nabla\left(\frac{\partial\eta}{\partial t}\right)\mathrm{d}\mathbf{x}
     \\
     &+ \alpha\int_\Omega\nabla\rho\cdot\nabla\left(\frac{\partial\eta}{\partial t}\right)\mathrm{d}\mathbf{x}
     +\alpha\int_\Omega\nabla\left(\frac{\partial\rho}{\partial t}\right)\cdot\nabla\eta \,\mathrm{d}\mathbf{x}
     \\
     &-\mu\int_\Omega\rho\,\frac{\partial\rho}{\partial t} \,\mathrm{d}\mathbf{x} -\int_\Omega\eta\,\frac{\partial\eta}{\partial t}\, \mathrm{d}\mathbf{x}
     -\omega\int_\Omega\rho\,\frac{\partial\eta}{\partial t} \,\mathrm{d}\mathbf{x} -\omega\int_\Omega\frac{\partial\rho}{\partial t}\,\eta \,\mathrm{d}\mathbf{x}
     \\
     =&\,\int_\Omega\rho\,\nabla\left(\kappa\Delta\rho +\alpha\Delta\eta+\mu\rho+\omega\eta\right)\cdot\nabla\left(\frac{\delta\mathcal{F}_2}{\delta\rho}\right)\mathrm{d}\mathbf{x}
     \\
     &+\int_\Omega\eta\,\nabla\left(\alpha\Delta\rho +\Delta\eta+\omega\rho+\eta\right)\cdot\nabla\left(\frac{\delta\mathcal{F}_2}{\delta\eta}\right)\mathrm{d}\mathbf{x}
     \\
     =&\,-\int_\Omega\rho\left|\nabla\frac{\delta\mathcal{F}_2}{\delta\rho}\right|^2\mathrm{d}\mathbf{x}-\int_\Omega\eta\left|\nabla\frac{\delta\mathcal{F}_2}{\delta\eta}\right|^2\mathrm{d}\mathbf{x}\leq 0.
\end{align*}
Again, we used the boundary conditions on $\rho$ and $\eta$ given by Eq. ~\eqref{eq:boundary_conditions}.

\subsection{Numerical simulations for Steinberg experiments in one dimension}

Here we study numerically whether the local model is able to reproduce the four different patterns observed in Steinberg experiments (Figure \ref{fig:DAH_twospecies}), namely: (i) mixing; (ii) engulfment; (iii) partial engulfment; (iv) sorting. Animated movies of the simulations in this section are available in~\cite{figshare}. To do so, we must determine which parameter ranges correspond to each of the observed patterns. This is simpler when we assume a particular shape for the potentials $\varphi_{ij}$. Let us then assume that these only differ by constants, i.e. $\varphi_{ij} = K_{ij}\varphi$, for constants $K_{ij}\geq 0$ satisfying $K_{12} = K_{21}$ and a given potential $\varphi$. This is actually the case in previously used nonlocal models~\cite{ArmstrongPainterSherratt,carrillo2018zoology,CarrilloMurakawaCellAdhesion}, where the constants $K_{ij}$ give the adhesive strengths of the two cell populations. Under these assumptions, the model parameters are directly related to the moments of $\varphi$ and the constants $\epsilon$ and $K_{ij}$:
\begin{equation*}
    \kappa = \frac{K_{11}}{K_{22}}\,,\quad\alpha = \frac{K_{12}}{K_{22}}\,,\quad\mu =\left( \frac{M_0-\epsilon/K_{11}}{M_0-\epsilon/K_{22}}\right)\kappa,\quad\omega =\left( \frac{M_0-\epsilon/K_{12}}{M_0-\epsilon/K_{22}}\right)\alpha,
   \end{equation*}
   where $M_0$ is the volume of the potential $\varphi$.
    Note then that $\kappa$ and $\alpha$ can be interpreted as relative adhesion strengths, as mentioned earlier. However, $\mu$ and $\omega$ are not only related to cell-cell adhesion but also to the strength of local repulsion due to volume exclusion.

    \begin{figure}
     \centerline{\includegraphics[width = \textwidth]{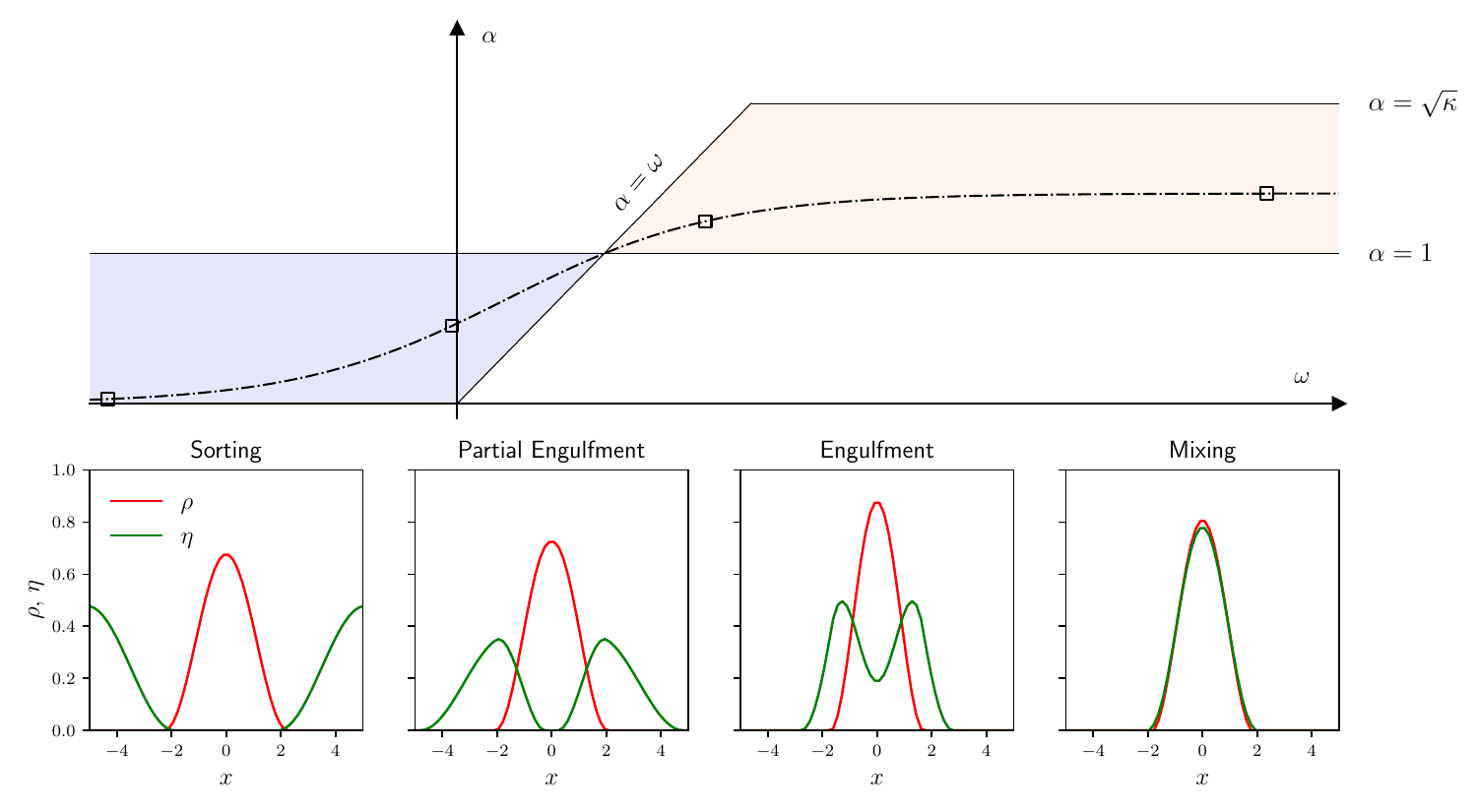}}
     \caption{Understanding the impact of changing model parameters. Imposing that $\eta$ is the less cohesive population implies $\mu>\kappa>1$ as discussed in the text. We focus then  on the cross-interactions. Parameter ranges for $\alpha$ and $\omega$ shown above: the blue-shaded region represents the weak cross-adhesion regime, while the red-shaded region corresponds to the case of strong cross-adhesion. Below we plot the numerically found steady states for the parameter values given by the square points: sorting, $\omega = -2.38,\,\alpha =0.03$; partial engulfment, $\omega = -0.04,\,\alpha =0.52$; engulfment, $\omega = 1.69,\,\alpha = 1.21$; mixing, $\omega = 5.51,\,\alpha =1.40$. In every case $\kappa = 2$ and $\mu = 4$. We observe the different patterns seen in the Steinberg experiments and the transition from sorting to mixing as we increase the cross-adhesion, agreeing with the model interpretation. Numerical simulations performed on a domain of length $L = 5$ and $\Delta x = 0.2,\,\Delta t = 0.01$ with periodic boundary conditions and initial condition $\rho(x,0) = \eta(x,0) = \mathds{1}_{|x|<1.5}/2$. See \textup{\cite{figshare}} for an animated movie with the stationary states corresponding to each point in the dashed line.}
     \label{fig:4patterns}
 \end{figure}

    We also assume without loss of generality that $\eta$ is the less cohesive population and hence $K_{22}<K_{11}$. Then according to these expressions we have $\mu>\kappa>1$. Parameter values outside of this range are also valid but  interpreting the model in such cases becomes more challenging. For the cross-interaction parameters $\alpha$ and $\omega$, one must distinguish two regimes depending on $K_{12}$, i.e. the strength of the cross-adhesion:
    \begin{enumerate}
        \item \text{Weak cross-adhesion} ($K_{12}<K_{22}$). In this case, we have $\omega < \alpha < 1$, as shown by the blue-shaded region in Figure \ref{fig:4patterns}. As the cross-adhesion strength decreases, $\omega$ becomes more negative, and $\alpha$ decreases as well.

       \item \text{Strong cross-adhesion} ($K_{12}>K_{22}$). Conversely, when the cross-adhesion is stronger than the self-adhesion of the second cell type, we have $\omega > \alpha > 1$, as depicted in the red-shaded region in Figure \ref{fig:4patterns}. In this case, both $\alpha$ and $\omega$ increase with the cross-adhesion strength. Additionally, the quotient $\omega/\alpha$ increases with $K_{12}$, meaning that in the limit of strong cross-adhesion, we should expect $\omega \gg \alpha$.

    \end{enumerate}

    Given the gradient flow structure of the local model in Eqs.~\eqref{eq:local_2species}, one could also understand these regimes by looking at the free energy $\mathcal{F}_2[\rho,\eta]$. We focus on the interaction terms in Eq.~\eqref{eq:energy2}. In particular, it becomes evident that whenever $\omega<0$, then in order to minimize the energy, both species will tend to separate so that the value of the product $\rho\eta$ is small. On the other hand, when $\omega>0$, the two cell types will be attracted to each other, trying to maximize the value of $\rho\eta$.

  We now simulate Eqs.~\eqref{eq:local_2species} having in mind the above considerations. We start with numerical simulations in small domains and periodic boundary conditions, as shown in Figure \ref{fig:4patterns}. The figure suggests that our intuition of the model was correct, since the described regimes are able to replicate the four patterns observed in the Steinberg experiments. As predicted by the differential adhesion hypothesis, we observe that the two cell populations tend to separate when the cross-adhesion is weak. On the other hand, if the cross-adhesion is larger, the more cohesive population $\rho$ gets engulfed inside $\eta$, and eventually the cross-adhesion is strong enough to drive mixing of the two.

  \begin{figure}[tp!]
    \centerline{\includegraphics[width = \textwidth]{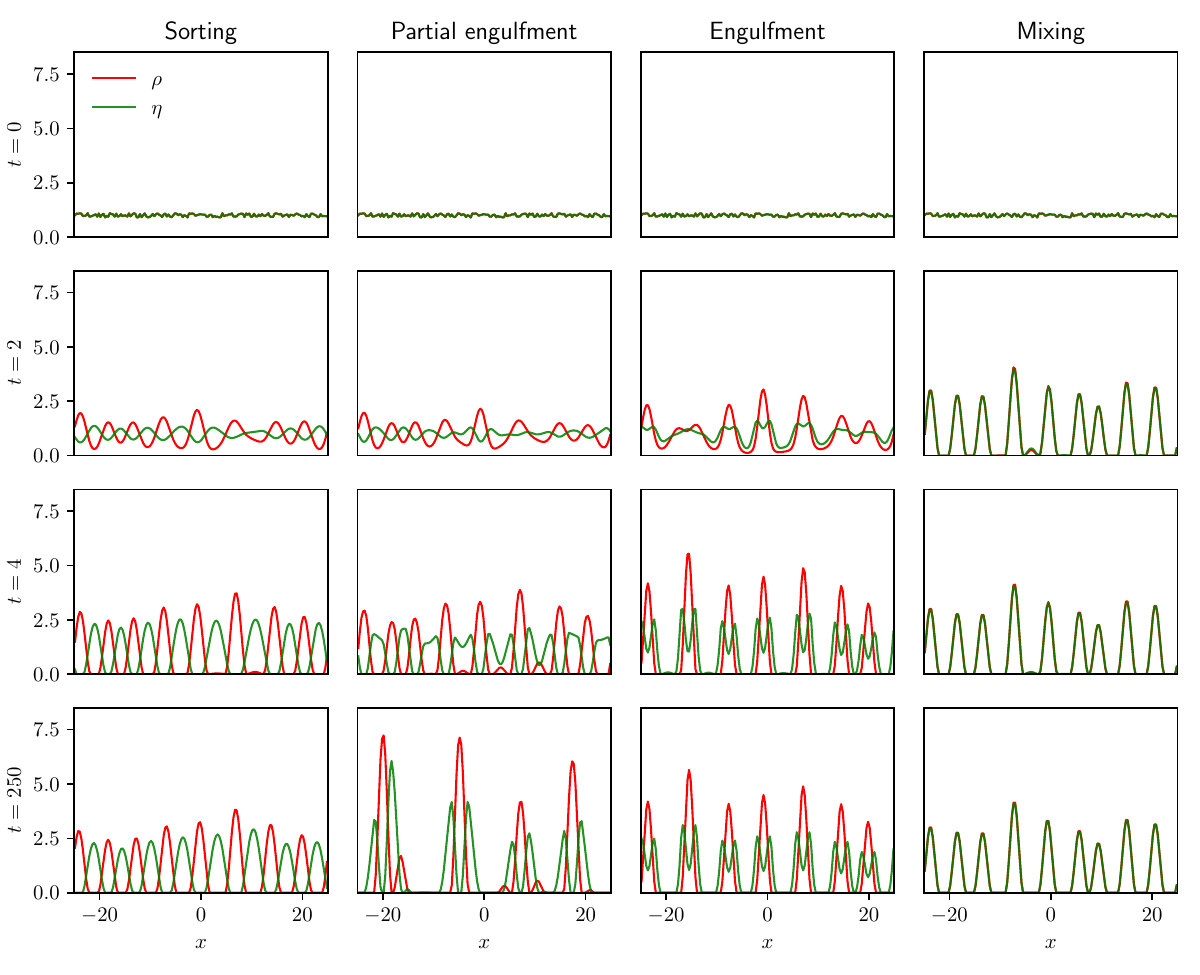}}
    \caption{Solutions of the local model using model parameters related to the Steinberg experiments. Each column represents the solution with the same set of parameters and at different times. Sorting, $\alpha = 0 ,\,\omega =-1$; partial engulfment, $\alpha = 0.8 ,\,\omega = 0.2$; engulfment, $\alpha = 1.3 ,\,\omega =2$; mixing, $\alpha = 1.4 ,\,\omega =6$. In every case $\kappa = 2$ and $\mu = 4$ and also  $L = 25,\,\Delta x = 0.2,\, \Delta t =0.01$. See \textup{\cite{figshare}} for animated movies.}
    \label{Fig:2species_1d}
\end{figure}

  The same patterns emerge in numerical simulations on larger domains. Here we choose parameter values corresponding to the two regimes that we described above, and show the solutions at different times in Figure \ref{Fig:2species_1d}. When the cross-adhesion is non-zero ($\alpha\neq 0$), steady states are composed of multiple compactly supported blobs or bumps. For this local model it is possible to find analytically the exact shape of each one of these bumps, given their individual masses. This is an advantage with respect to nonlocal models, where analytical solutions are only available for specific types of  potentials~\cite{carrillo2018zoology}. We refer to~\cite{falco2024local} for more details. Note, however, that predicting the final mass of each of the bumps is difficult.

\begin{figure}
    \centering
    \includegraphics[width=0.97\linewidth]{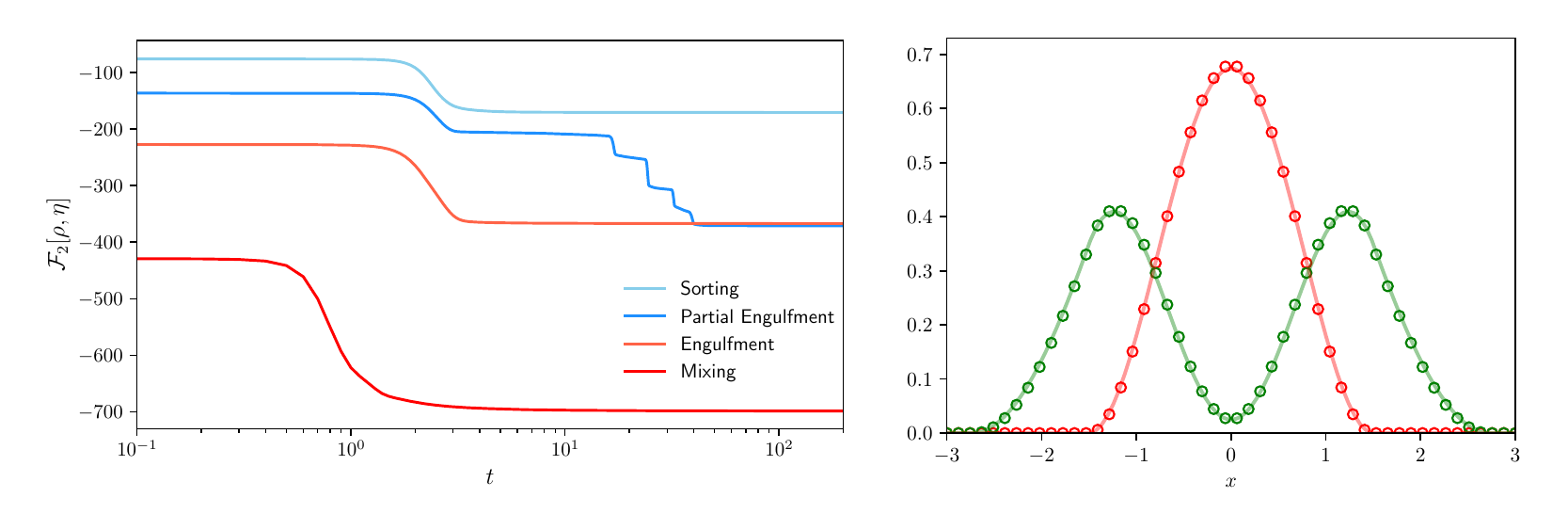}
    \caption{(left) Energy decay given by Eq.~\eqref{eq:energy2} for the numerical solutions in Figure \ref{Fig:2species_1d}. Solutions corresponding to the weak and strong cross-adhesion regimes represented in blue and red, respectively. In general, the stronger the cross-adhesion, the faster the decay of $\mathcal{F}_2[\rho,\eta]$. (right) Numerical solution of the two-species model with initial condition $\rho(x,0) = \eta(x,0) = \mathds{1}_{|x|<1}/2$. Solutions are shown at $t = 25$ (solid line) and the corresponding analytical stationary solutions are also plotted (dots, see~\cite{falco2024local} for details on how to calculate these). The analytical and numerical stationary solutions agree perfectly. Simulation parameters: $\kappa = 2,\,\alpha = 1.3,\,\mu =4 ,\,\omega =1.8,\,m = 1,\, L = 3,\, \Delta x = 0.1,\, \Delta t = 10^{-2}$.}
    \label{fig: 5}
\end{figure}

Observe too that solutions in the weak cross-adhesion regime --- corresponding to the sorting and partial engulfment patterns --- show very similar behavior for early times (see Figures \ref{Fig:2species_1d} and \ref{fig: 5}). When $\alpha = 0$ then both cell species tend to separate, converging to more or less sharply segregated solutions --- which does not happen in previous nonlocal models that consider linear diffusion~\cite{ArmstrongPainterSherratt}. However, when cross-adhesion is small but strictly positive, the two populations move away from each other at early times and later organize themselves to form different aggregates, composed by different coexistence regions. This kind of metastability has also been observed before in Cahn-Hilliard type systems~\cite{barrett2001fully,celora2021dynamics, falco2024local}.

 \subsection{Numerical simulations for Steinberg experiments in two dimensions}

\begin{figure}[tp!]
    \centerline{\includegraphics[width = \textwidth]{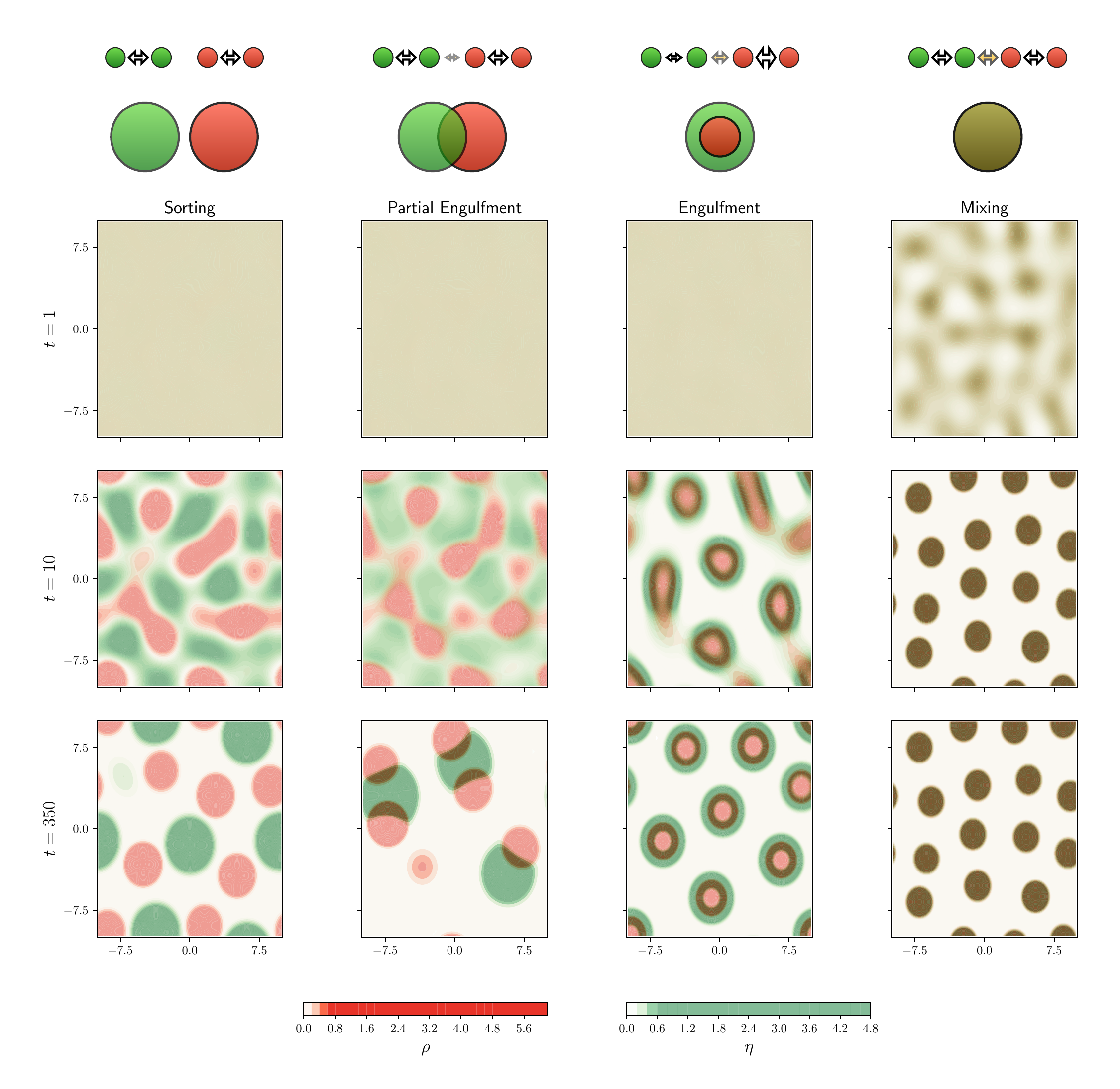}}
    \caption{Numerical solutions of the local model using model parameters related to Steinberg experiments. Each column represents the solution with the same set of parameters and at different times. Sorting, $\alpha = 0 ,\,\omega =-1$; partial engulfment, $\alpha = 0.5 ,\,\omega = -0.02$; engulfment, $\alpha = 1.3 ,\,\omega =2$; mixing, $\alpha = 1.4 ,\,\omega =8$. In every case $\kappa = 2$ and $\mu = 4$ and also  $L = 10,\,\Delta x = 0.2,\, \Delta t =0.001$. The initial condition is the same for every experiment, $\rho(x,0), \,\eta(x,0) = 0.3$ plus a small perturbation. See \textup{\cite{figshare}} for animated movies.}
    \label{fig:2D_DAH}
\end{figure}

Here we explore the model in two dimensions numerically, performing the same type of experiments as in the one-dimensional case, which we show in Figure \ref{fig:2D_DAH}. By choosing appropriate parameters, as explained in the previous sections, we can again recover the four patterns seen in the Steinberg experiments. The model dynamics are in general similar to the one-dimensional case, with the strong cross-adhesion regime showing a faster decay to the stationary solution. Note that although the final configurations are very different for the chosen parameters, in the weak cross-adhesion regime solutions show very similar patterns for early times. In this case, the model shows two separated timescales with solutions showing large differences only after the first one, as is also the case in one spatial dimension (see Figure \ref{Fig:2species_1d}). In the limit of vanishing cross-adhesion we recover the typical cell sorting pattern with sharp segregation of the two species, which is only observed in models that account for population pressure. Note also that this pattern is accentuated with respect to the one-dimensional case.

Finally, we have implemented the two-species version of the model in VisualPDE~\cite{visualPDE}, offering an interactive platform that allows the reader to easily explore how different patterns depend on the model parameters. This implementation may introduce numerical errors, and we advise using the finite-volume scheme in~\cite{falco2024local} for a more accurate approximation. Links to simulations for all cases discussed here --- \href{https://visualpde.com/sim/?options=N4IgRghiBcIM4FsD2SAuALCATLIA04AbjCAHQCM+IAxiVgJaECmATgOb0B2bV1AIjFQsArkwJYAqhJIBaBMIBUCCAA8AFMLwAGAJRVJANVlIETNhCWqN2veIkB1WeUvrNu-QemwZJsxeXqhDYeRt7OAWpB7uIGAIpOLpHB4vZeIADWEAAOWVApoSAQADZZmPr2jrBa+rFpxaV5IFixBZTicCQATNXiqDBapFpDWuQE1GltNK1U6ST1mAAECwC8S1pLXEsA2lp4S+RaALoA3Jk5EEurC50bnNu7+0fHvuaXSzLkt9sy3XvXT-I3gsACxfBY7P4HE6oAD6cCBEQgYDgankOj+SJRLwgOh0xyoyi6BPosk6VE4UGgnQInBgwJpABUYAdqSAsiQAMLCOCoExULIsJBgJgAMWEtNgwn5guFABkmNwMMyAKwEFhpHogFgFTUsSogTVwJioADKTCYuGgQlEBDg7KqpE6yqocAAjoIRGJ4FKrZ6CKh6KY4NRikwOr6bSAA0GQ0UmDDKLBYR1-YGw7H42Sk3CqMRoFtQBAWIKAO4cpBFJDCFgwABsyuVAGZVYVi0gS3wFXB6KgAJ79Ugtoul+WK9AAWVUJEmw-bJozc2EvKos5LAA0SLKYSoAPQ3GQLFQrtslgCam5hvb3CwP-YIECXSDNqHLlerACUINwmDAAGbFI0xggUwWAgBl0GNSlG12GhgNYCAAAV0BJaBoKAkCIAALRQBABwADjGCsqxYZR7SaRhWA4bheCI6tIBraB-yKQCaCQTheWrcMmJY6g2I4lhX2I-pCPY4iAFEsm7SsJQGEYRP4gA5YRcOgFtqG5XkEBNat-2oH9GIAr0mAQMAkDgLjDIIYzTPMgBBEz6AVPpoAGFtrLMuA+HoX9f25fSBlrKyTI8pCUPIQcAHYtAIkB3PMk1kM4ftoAOIKbLgE1kDQdBODDcMBlZOKMqyJh1KKIsB0K4LzPAyCBwivDlXpEBfyKegcgtQTq1Iv9LJANrcvsegsAwcdhCKZkCACLrtWKUQiSmrgZoMOb9JAH4qCQZgWHK5LuK9LbWF2mbhJAQ6dogXsJKktiByGFtzt2sSVAFadNu23bZS4JghpGidxsmtlK1QBlexKkgsnK3LpSFLgeAM5ivQFIVRXFEgpQIZG5QVNglRSlssaYUHwdgLhUDMUCJsxmUmA3KoYeFc96dtHSID0sUJQNKgMBYc0+AACSYeg2HQU0Fxc0gIoIZhqF5FgRUcopLX2ggS0wEGkAQ4H0ZkYgaTgt6AF9DlVy0QAxkASzI4RdZdQMGR7ONNyQWMFg5JgiiKGQ9M9hZsAg7s2IWZAsA9m9rgWO1Ssc+E1DgJAWADbgFhDI09ENoA}{sorting}, \href{https://visualpde.com/sim/?options=N4IgRghiBcIM4FsD2SAuALCATLIA04AbjCAHQCM+IAxiVgJaECmATgOb0B2bV1AIjFQsArkwJYAqhJIBaBMIBUCCAA8AFMLwAGAJRVJANVlIETNhCWqN2veIkB1WeUvrNu-QemwZJsxeXqhDYeRt7OAWpB7uIGAIpOLpHB4vZeIADWEAAOWVApoSAQADZZmPr2jrBa+rFpxaV5IFixBZTicCQATNXiqDBapFpDWuQE1GltNK1U6ST1mAAECwC8SwMArEtcSwDaWnhL5FoAugDcmTkQS6sLnVucu-uHJ6e+5tdLMgNadwvbCzsZN0DrcXvIPgsACz3R4go5nVAAfTgEIiEDAcDU8h0IPRmLeEB0OlOVGUXVJ9FknSonCg0E6BE4MEhjIAKjAAOzkdYELIkADCwjgqBMVCyLCQYCYADFhEzYMIxRKpQAZJjcDAwbkEFhpHogFgFfUsSogfVwJioADKTCYuGgQlEBDgfKqpE66yocAAjoIRGJ4IqHf6CKh6KY4NRikwOsGnSAwxGo0UmIjKLAkR1Q+GY8nU9SM8iqMRoDtQBAWBKAO78pBFJDCFgwABs63WAGYeYVK0gq3x1XB6KgAJ79UhdivVtUa9AAWVUJEmk97Vrzc2EIqoy6rAA0SCrESoAPR3GQLFRbntVgCa+8Rw5PCzPo4IEA3SBtqFr9cbACUINwTAwAAZsUFpjBApgsBArLoJadLtvsNCQawEAAAroJS0CIRBUEQAAWigCBjgAHGMdYNiwyiuk0jCsBw3C8BRjaQE20CgUU4E0EgnAio2sYcVx1A8XxLDfpR-TkbxlEAKJZIO9byt8ozcdJjYAHLCMR0BdtQQoiggVqNqB1BAexYEBkwCBgEgcACRZBBWTZdkAILWfQ6p9NAGyOdZtlwHw9DAcBQpmQMza+c5cAYVh5DjhyWhkSATn+VamGcKO0BHJFqXIGg6CcDGsYDAyyV+XZVpZEwelFBWY6lSldmwfBY4ciR6wsiAwFFPQOR2uJjbUSBDkgD1hX2PQWAYLOwhFFqBABANhrFKI5ILVwS0GCtZkgECVBIMwLC1ZlgkBgdrDHUtkkgOdR0QMOckKTxY5DF2t3HTJKjiou+2HcdKpcEwE1TXOs3zSAWT1qgrLDlVJCQwBQG8sqXA8OZnEBuKkoynKJCKsj2PTmwmpZV2WNSjDcOwFwqBmNBc0E1Ke5VEq2O3izzrGRApmyvKZpUBgLC2nwAASTD0Gw6DWmu3mkByBDMNQIosNKHlFPap0EFWmDQ0gaFQ3jMjEIyKE-QAvscWv2iA+MgFWNHCEbXrhqyQ4pvuSDJgs-JMEURQyKZfsLNgcGDjxCzIFgvtPrcCwutVHkomouQsGGxQLOqbCzcBpi8XoZtAA}{partial engulfment}, \href{https://visualpde.com/sim/?options=N4IgRghiBcIM4FsD2SAuALCATLIA04AbjCAHQCM+IAxiVgJaECmATgOb0B2bV1AIjFQsArkwJYAqhJIBaBMIBUCCAA8AFMLwAGAJRVJANVlIETNhCWqN2veIkB1WeUvrNu-QemwZJsxeXqhDYeRt7OAWpB7uIGAIpOLpHB4vZeIADWEAAOWVApoSAQADZZmPr2jrBa+rFpxaV5IFixBZTicCQATNXiqDBapFpDWuQE1GltNK1U6ST1mAAECwC8S+SkAMxLXEsA2lp4a1oAugDcmTkQS6sLnducewdHZ77m10t3CzsLuzLdh7cTqd5O8FgAWe6PAHkIGoAD6cFBEQgYDgankOgBKLRrwgOh0pyoyi6RPosk6VE4UGgnQInBgYLpABUYAB2AAcrIIWRIAGFhHBUCYqFkWEgwEwAGLCemwYQisUSgAyTG4GBg5AArAQWGkeiAWAV9SxKiB9XAmKgAMpMJi4aBCUQEOA8qqkTqaqhwACOghEYng8od-oIqHopjg1GKTA6wadIDDEajRSYcMosHhHVD4ZjydTFIzCKoxGgu1AEBYYoA7rykEUkMIWDAAGyazUbbWFStIKt8VVweioACe-VInYr1ZVavQAFlVCRJhOe1a83NhEKqEuqwANEhKuEqAD0dxkCxUm+7VYAmnu4UPjwtTyOCBB10gbaha-XGwAlCDcJgYAAM2KC0xggUwWAgJl0EtakNgOGgINYCAAAV0DJaAEPAyCIAALRQBBR3ZMY6wbFhlFdJpGFYDhuF4MjG0gJtoBAoowJoJBOCFRtYzYjjqC4niWC-cj+lI7jyIAUSyAd61lAYRgk4SADlhCI6BO2oAUhQQK1GxA6hANY0CAyYBAwCQOA+NMghzMs6yAEELPoVU+mgAZO3sqy4D4eggKAgVjIGZs7Isnz0Mw9ZNVZLQSJAbzrKtDDOBHaAYTChy4CtZA0HQTgY1jAZaQS8KkqyJhtKKCtRxKxK4BguDRw5TVGRAICinoHI7VExtKOA2yQE6gr7HoLAMBnYQig1AgAl6w1ilEElZq4eaDEW4yQD+KgkGYFhqrS-iA121gDvm8SQBO-aICHGS5K40chk7K6DqklRRQXHa9oOpUuCYUbxtnKaZpALJ61QJkhwqkgwf-QDuUVLgeBM9iA1FcUpRlEh5QRjGpzYdV0s7dGJUh6HYC4VAzCg6bcYlXcqgVDGb0Z50DIgIzpVlM0qAwFhbT4AAJJh6DYdBrVXDzSC5EBmGoIUWElVyintI6CCrTAIaQVDwexmRiDpZDPoAX2OdX7RAHGQCrKjhH1r1wyZQcUz3JBkwWXkmCKIoZCM72FmwWCBy4hZkCwL3H1uBYXUq1zETUVU2CmoDTG4vRjaAA}{engulfment}, \href{https://visualpde.com/sim/?options=N4IgRghiBcIM4FsD2SAuALCATLIA04AbjCAHQCM+IAxidegKYICW1EANgOIBODDAdlWoARGKm4BXBgSwBVWSQC0CCQCoEEAB4AKCXgAMASipyAakqQIGAcwjqtug8ZmyA6kvL2deoydMLYRUsbOw0dQic-c0DPMO0I3xlTAEUPL3jImVcAkABrCAAHAqgs6JAOAswTV3dYfRNknIqqmWSyyhk4EgAmeplUGH1SfRH9cgJqHI6adqpckmaIAAIlgF4V8lIAFhXmfhWAbX08Df0AXQBufKLltZXu3f2lo5Ol8nOL4NsV9aWADkeh0UvVevUuKh+Kx2Sz2h2Op0uqAA+nBIUs4hAwHBtCpDK9MdivhBDIYLlQNCQAMzk5gkPogfhQaDdAiCaBbVkAFRg7xGBAKJAAwhI4KhLFQCtwkGAGAAxCRskASCVSmUAGQE1gwPIArARuDl6dwykbaiB6XAGKgAMp8XDQcRSAhwAV1UjdHVUOAARzEkmk8GVDv9BFQzCscDY7AYXWDTpAYYjUYYSMosGRXVD4ZjyaR3RIGaoxGgB1AEG4UoA7oKkOwkBJuDAAGw6nWUvXlCtISvCARwZioACeg1IHfLVY1-C16AAsloSNNx93rcmFhIxVQl5WABokNVIzQAegeiiWmk3XcrAE090jB8elqfhwQIOukLbUDW6w2AEoQKcMDAABmHCWhMEBWNwECcowqBMpSxw0BBDBQQACugtLQAh4GQRAABaKAICOfwTLW9bcBoro0IwLBRjwfCCKR37cJAjbQCB7BgTQSD8GKDaxhxXHUDxfHcF+5GDKRvHkQAogU-Z1myQxjFJokAHISER0AdtQIpigg1oNiB1CAexoEBkwYBIHAAnmQQlnWXAACCCBgMwAgDNAQwdg5NnCMwQFASKplDE29muY56GYZsOoAOz6CRIC+XA1oYfww7QO84VWTZ1rIGg6D8DGsZDCySURblBQMLp7DliOZXJTBVpMkMsV-DqHIgEB7DMEUDBYOJDaUcBdkgD1RWuMwWAYDOEjsDyBBhINxocFIVI0vwy2mKtpnmlQSCEChtUZYJAYHUdECDstkkgOd3DHXJCk8SOIwdndD2aJKC77Yd92XWqewMJN02znNC0gAUdaoJyg5VSQkP-oB-Kqns1gjZxAaStKcoKiQyrI9jk7TrqBMyjDcOwHsqA2FB82kwwu51Cq2M3kzzpGRAJnyoq9IYLwDDCAAEgwzDWOgNqrl5pCxQQh3UGK3Cyu57D2qdBCVpg0NIKhUMkMQrLId9AC+Zzq-aSpeuGnIDtGe5IFGSyCgw7DsIoJku0s2CMP2PHokgWDO4+SwPC61XuaiOLMJoqPGEbQA}{mixing} --- are available for further exploration.

\section{Conclusion, open problems, and outlook}
We have presented a local continuum model of aggregation-diffusion in the form of a system of thin-film equations, and used this framework to describe adhesion-based pattern formation. The idea of describing tissues using fluid-like properties has been recurrent. In fact, Steinberg first employed this analogy in developing the differential adhesion hypothesis. However, the model proposed here differs from other descriptions based on the fluid analogy~\cite{AlertTrepat}, as it can be directly related to aggregation-diffusion equations. The local model has physically interpretable parameters and successfully explains the patterns observed in experiments, as well as those predicted by the differential adhesion hypothesis. This had only been achieved in the continuum setting with nonlocal models~\cite{falco2024local}.

There are also more modern views on the differential adhesion hypothesis, such as the differential interfacial tension hypothesis~\cite{BrodlandDITH,reviewCellCellAdhesion}. This hypothesis suggests that tissue surface tension is determined not only by adhesion bonds between cells but also by cortical tension~\cite{amack2012knowing,youssef2011quantification}. Some modeling efforts have aimed to incorporate these factors, showing that when cell-cell adhesion is the dominant interaction, the differential adhesion hypothesis successfully predicts tissue behavior~\cite{ManningFotySteinbergSchoetz}. However, when cortical tension is stronger, the differential adhesion hypothesis may no longer be sufficient, indicating that in this regime cell shape plays an important role. While this is an important consideration, our model here focuses on the adhesion-based regime, where both the differential adhesion hypothesis and the particle-based approximation hold.

The approach taken here is based on previous studies of aggregation-diffusion models, where the nonlocal terms are approximated by a series of terms including higher-order derivatives of the densities~\cite{BernoffTopazCH,DelgadinoCH} --- these consider a porous-medium type repulsion with exponent three, instead of the exponent two considered here. Following a similar approach, energy minimizers and linear stability for multi-species systems have also been explored~\cite{EllefsenRodriguez}. Bernoff and Topaz~\cite{BernoffTopazCH} show that energy minimizers of the local and nonlocal models are in good agreement and share similar qualitative properties in the limit of large populations, far from aggregation boundaries. In our case, the choice of diffusion is different so this result may not directly apply. However, recent work~\cite{carrillo2024degenerate, elbar2022degenerate} demonstrates that, in a similar setting with unit volume and compactly supported potential, the nonlocal model converges to its local approximation as the scaling parameter tends to zero ($a \rightarrow 0$). Recent numerical investigations by Buttenschön et al.~\cite{buttenschon2024cells} suggest more significant differences between the local and nonlocal models when the interaction potential has non-compact support.

In our recent work, we established the existence theory for the local model in both the one-species and two-species cases, as well as for a more general second-order aggregation term~\cite{carrillo2024competing, falco2024local}. However, several analytical challenges remain unresolved. Uniqueness is an open problem, as the functionals involved are not convex. The existence of free energy minimizers in the entire space is also missing, as we do not have a method to control the escape of mass at infinity. Additionally, long-time asymptotics remain an open question in all global existence cases. We refer to~\cite{carrillo2024competing} for further discussion on analytical open problems.

The local model is, in principle, less complex and more analytically tractable than its nonlocal counterpart. Thus, we believe it could offer significant advantages for applications. Although solving numerically fourth-order equations can be challenging, there is a simplification in the numerical scheme complexity when one approximates convolutions with local operators~\cite{nonlocalNumericalSchemeRafaMarkus,nonlocalSchemeCarrilloYanghong}. From an applications perspective, it still needs to be tested whether the model can fully capture the dynamics of co-culture experiments involving multiple cell populations with different adhesive properties. While we have confirmed that the model can capture steady states and these are consistent with the differential adhesion hypothesis, further exploration is needed to assess the extent to which it can be applied to these settings. To this end, one could estimate the different parameters in the model using computational inference techniques \cite{falco2023quantifying}. This is now more feasible with the local model, as we only need to infer the parameters directly, rather than the interaction potentials required in the nonlocal model. We expect that the use of local models will open up further developments in the inference of aggregation-diffusion phenomena across various fields.

\section*{Acknowledgments}The third author was supported by the Advanced Grant Nonlocal-CPD (Nonlocal PDEs for Complex Particle Dynamics: Phase Transitions, Patterns and Synchronization) of the European Research Council Executive Agency (ERC) under the European Union's Horizon 2020 research and innovation programme (grant agreement No.~883363).
The third author  was also partially supported by EPSRC grants EP/T022132/1 and EP/V051121/1. The first author  acknowledges support of a fellowship from ``la Caixa'' Foundation (ID 100010434) with code LCF/BQ/EU21/11890128.

\end{document}